\documentclass[aps,prc,twocolumn,tightenlines,showpacs,superscriptaddress,nofootinbib,a4paper]{revtex4}
\usepackage{graphicx,color}
\usepackage{braket}
\usepackage{amsmath}
\usepackage{amssymb,amsbsy}
\newcommand{\be}{\begin{eqnarray}}
\newcommand{\ee}{\end{eqnarray}}
\newcommand{\dd}{{\rm d}}

\newcommand{\LL}{{\cal L}}

\newlength{\figwidth}

\begin{document}
\setlength{\figwidth}{0.98\columnwidth}
\def\vec#1{{\boldsymbol{#1}}}

\title{Intermediate energy four-body breakup calculations for $^{22}$C}

\author{Y. Kucuk}
\altaffiliation{Present address: Akdeniz University, Antalya, Turkey}
\affiliation{Department of Physics, Giresun University, 28100-Giresun,
Turkey}
\affiliation{Department of Physics, Faculty of Engineering and
Physical Sciences, University of Surrey,
Guildford, Surrey GU2 7XH, United Kingdom}
\author{J.\,A. Tostevin}
\affiliation{Department of Physics, Faculty of Engineering and
Physical Sciences, University of Surrey, Guildford, Surrey GU2 7XH,
United Kingdom}

\begin{abstract}
The heaviest particle-bound carbon isotope, $^{22}$C, is thought to have a Borromean
three-body structure. We discuss and compare four-body, i.e. three-body projectile plus
target, reaction model calculations of reaction cross sections for such systems that
use the fast adiabatic approximation. These methods are efficient and well-suited for
quantitative analyses of reactions of neutron-rich nuclei with light target nuclei at
secondary beam energies of $\approx$300 MeV/nucleon, as are now becoming available.
We compare the predictions of the adiabatic model of the reaction both without and
when including the additional eikonal approximation that has been used extensively.
We show that the reaction cross section calculations have only limited sensitivity
to the size and structure of $^{22}$C and that the differences arising from use of
the eikonal approximation of the reaction mechanism are of a similar magnitude.
\end{abstract}

\date{\today}
\maketitle

\section{Introduction}

Reaction and interaction cross sections of light neutron- and proton-rich
projectiles with light target nuclei have been used extensively as an
inclusive observable with sensitivity to projectile size and binding.
Calculations of the elastic scattering, reaction and breakup observables
of weakly-bound two-neutron-halo nuclei, the ground states of which can be
modelled using three-body-model wave functions, is a four-body reaction
problem. As was shown in previous three- and four-body analyses of the
reaction cross sections for weakly-bound nuclei, the treatment of breakup
degrees of freedom is essential to obtain a quantitative description of the
reaction observables. Practical calculations that treat the few-body structure
of the projectile explicitly are now possible. For example, at high energies,
of order 800 MeV/nucleon, the eikonal (Glauber) theory has been employed
\cite{jak1,jak2} that provides an efficient approximate methodology.
Counter to ones intuition, the inclusion of strongly-coupled breakup
channels, that remove flux from the elastic channel, actually resulted in
calculated reaction cross sections that were smaller than those obtained
using simpler (no-breakup) one-body density-based models, which overestimate
the contribution to the cross sections from the dilute halo of one or more
weakly-bound neutrons \cite{jak1,jak2}. This is also expected to be the
case for reactions of the heaviest particle-bound carbon isotope, $^{22}$C,
for which first measurements are now possible, see e.g. \cite{Tanaka}. That
data set suggested an enhanced $^{22}$C reaction cross section on a proton
target at relatively low energy and the authors used a simplified reaction
model to infer a very large root-mean-squared (rms) matter radius for $^{22}$C.
Inclusive measurements of Coulomb dissociation of $^{22}$C on a heavy (Pb)
target \cite{Naka} and of fast nucleon removal on a light (C) target \cite{Kob}
have also recently been performed at RIKEN at energies in excess of 200
MeV/nucleon, also displaying enhanced reaction yields characteristic of weak
binding.

In this paper we exploit the coupled-channels adiabatic approach of Christley
{\em et al.} \cite{Christley} for (four-body) model calculations of reaction
cross sections, $\sigma_R$. We compare these calculations with the adiabatic
plus eikonal four-body dynamical approach \cite{jak1,jak2} that makes additional
approximations. Our primary objectives are twofold. (i) To quantify the
sensitivity of the $\sigma_R$ to the structure assumed for this weakly-bound
dripline (and Borromean) system and clarify the breakup channels of most
importance. Since the adiabatic method is computationally more efficient, a
knowledge of this convergence may offer insight into the likely model space(s)
needed for higher-energy calculations using e.g. the coupled discretized
continuum channels approach \cite{cd1,cd2}. (ii) To quantify the differences
between the $\sigma_R$ calculations that use and do not use the additional
eikonal approximation(s) to the adiabatic model. To do this we explore the
$^{22}$C nucleus ground-state wave function within a core plus two-valence
neutron ($^{20}$C+n+n) three-body framework \cite{Efaddy} and the reaction
dynamics using the four-body adiabatic model, see \cite{Christley}. The
converged results of this approach are compared with calculations that use
the same $^{22}$C wave functions and optical model interactions but which
make the additional eikonal approximation to the reaction dynamics of the
four-body problem: as have been used previously \cite{jak1,jak2}. This paper
supersedes the brief report of preliminary results of the present study of
Ref.\ \cite{Ruth}.

\section{Projectile three-body model}

The nucleus $^{22}$C and its neutron-unbound subsystem $^{21}$C remain
poorly understood. Both the two-neutron separation energy from $^{22}$C
and the ground-state energy of $^{21}$C are poorly determined. The 2003
mass evaluation \cite{Aud03} gives $S_{2n}$(22)= 0.4(8) MeV and $S_{1n}
$(21)=$-0.3(6)$ MeV, both with large uncertainties. A recent direct mass
measurement places a limit of $S_{2n}(22)=-0.140(460)$ MeV \cite{Gaud},
however, since $^{22}$C is known to be bound, this might be interpreted
as $S_{2n}(22)<0.32$ MeV \cite{Mos}. Irrespective of such details, $^{22}
$C($^{21}$C) is certainly bound(unbound) and thus $^{22}$C has a Borromean,
$^{20}$C+n+n three-body character and is very interesting structurally.
The shell-model suggests that this $N=$16 nucleus will, predominantly, be
described by a $\nu [1d_{5/2}]^6\,[2s_{1/2}]^2$ closed neutron sub-shell
configuration. The expectation is therefore that $^{22}$C will have an
extended, predominantly $s$-wave two-neutron-halo wave function leading
to large reaction, nuclear, and Coulomb dissociation cross sections in
its collisions with a target nucleus.

A very recent analysis of new measurements of (inclusive) neutron removal
reactions from the most neutron-rich carbon isotopes \cite{Kob}, also made
at the RIBF, RIKEN at 240 MeV per nucleon, is broadly consistent with this
shell-model picture. The $^{22}$C($-$1n) data suggest a large spectroscopic
factor (of $\approx$1.4) for $2s_{1/2}$ neutron removal to an unbound
$^{21}$C($1/2^+$) ground state, which subsequently decays by neutron emission
to $^{20}$C. These data are also consistent with (though not highly sensitive
to) the $S_{2n}$(22) and $S_{n}$(21) of the 2003 mass evaluation \cite{Aud03}
and the $S_{2n}$(22) limit of the recent direct mass measurement \cite{Gaud}.

This same set of measurements \cite{Kob} also identifies a non-negligible
$\nu [2s_{1/2}]^2$ component in the (isolated) $^{20}$C($0^+$) ground-state,
manifest as population of the $^{19}$C($1/2^+$) ground-state in the neutron
removal reaction from $^{20}$C projectiles. This complication and the finer
details of the structure of the $^{20}$C core within $^{22}$C will not be
explored further in this paper.

\subsection{Model wave function}

\begin{figure}[t]
\begin{center}
\includegraphics[width=0.7\figwidth]{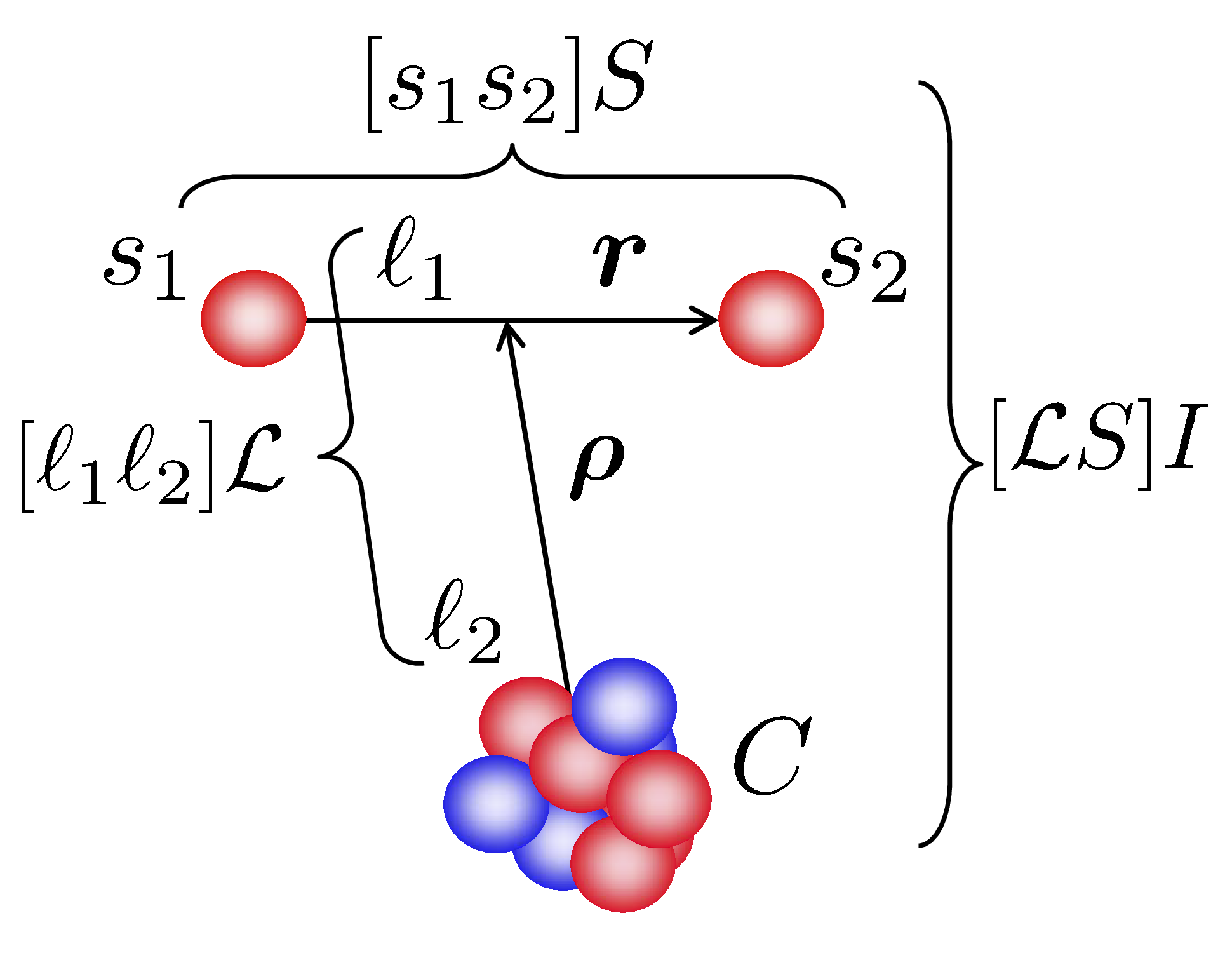}
\end{center}
\caption{\label{fig1} (Color online) Schematic representation of
the angular momentum decomposition and the angular momentum couplings (in the
Jacobi T-basis) used for the description of the $^{20}$C($0^+$) core+n+n
Borromean three-body projectile in its ground and (continuum) excited states.
In the case of the $^{22}$C ground state, with $I^\pi=0^+$, ${\cal L}=S$ and
$[s_1 s_2]S=0,1$ and, with $[\ell_1 \ell_2] {\cal L}$, ${\cal L}^\pi= 0^+$
or $1^+$.}
\end{figure}

Here we treat $^{22}$C using the three-body model, shown schematically in
Figure \ref{fig1}, as a $^{20}$C($0^+$) core+n+n system. Related $^{22}$C
structure studies can be found in Refs.\ \cite{japan,yamas}, on which we
comment. The core is assumed to have a filled $\nu [1d_{5/2}]^6$ sub-shell.
The n-n and nn-core relative orbital angular momenta are $\ell_1$ and
$\ell_2$ in our chosen (T-basis) Jacobi coordinate set $\vec{r}$ and
$\vec{\rho}$, respectively.

The projectile's ground state wave function (and the structure of the wave
function in the breakup channel with total angular momentum $IM_I$) is written,
in general, as a sum of individual angular momentum components
\begin{equation}
\Phi_{I M_I}(\vec{\rho},\vec{r}) =\sum_{\ell_1 \ell_2 \LL S} \left[ [\ell_1
\otimes\ell_2]_{\LL}\otimes S\right]_{I M_I}\,\frac{U_{[[\ell_1 \ell_2]\LL S]
I} (\rho,r)}{\rho r}, \label{ground}
\end{equation}
only a small number of which are expected to have significant probabilities,
as discussed below. The two-neutron configurations are thus $^{(2S+1)}\ell_1$.
For the $^{22}$C ground state $I^\pi=0^+$ and thus ${\cal L}=S$ and, given that
$[\ell_1 \ell_2] {\cal L}$, then ${\cal L}^\pi = 0^+$ or $1^+$.

\subsection{Three-body model parameters and results\label{3bwf}}

To solve for the three-body wave function we use the Gogny, Pires and De Tourreil
({\sc gpt}) interaction \cite{gpt} for $V_{nn}$. The neutron-$^{20}$C core
interactions $V^\ell_{n20}$ are described by Woods-Saxon plus spin-orbit
interactions. For all $\ell_j$-states we use radius and diffuseness parameters
1.25 fm and 0.65 fm, respectively, and a spin-orbit strength $V_{\rm so}$=6.3 MeV.
This (derivative Woods-Saxon) neutron-$^{20}$C spin-orbit potential is defined
according to the convention of Becchetti and Greenlees \cite{bandg}.

The depth of the central $d$-wave neutron-core interaction, $V^{\ell=2}_0$=42.0
MeV, was chosen to bind the neutron+$^{20}$C $1d_{5/2}$ state (by 2.3 MeV)
while the $1d_{3/2}$ state is unbound by 1.9 MeV. This fixed potential was used
for all $n\ell_j$ neutron+$^{20}$C configurations other than the $s$-wave
states. For the $s$-states this depth is too strong, binding the $2s_{1/2}$
state. The $s$-state well depth, $V^{\ell=0}_0$, was thus adjusted (reduced)
so that the n+$^{20}$C $2s_{1/2}$ state is unbound, as expected empirically.
This two-body $s$-state potential depth and the strength, $V_{3B}$, of an
added attractive central hyperradial three-body force,
\begin{equation}
V_{3B}(\varrho)=-V_{3B}/(1+[\varrho/5]^3)\ ,\label{3bf}
\end{equation}
were then used as parameters to generate bound $^{22}$C three-body wave functions
with a range of three-body energies, allowed by the uncertainty on the evaluated
$^{22}$C two-nucleon separation energy. We use $V_{3B}=1.6$ MeV. Each wave function
is thus characterised by (a) the position of the n+$^{20}$C $s_{1/2}$ virtual
state pole (located using a complex $k$-plane S-matrix search \cite{poler})
and its associated scattering length $a_0$ and (b) its bound three-body energy
eigenvalue, $E_{3B}=-S_{2n}$. Even in the absence of the added three-body force
$V_{3B}$, $^{22}$C is found to be bound with the potential set used, provided
that $a_0 \leq -46$ fm. With $V_{3B}=0$ and the largest $a_0$ value, model k5
in Table \ref{tab1}, $S_{2n}(22)=134$ keV.

\begin{table*}[ht]
\caption{\label{table1} Three-body model wave functions for the ground state
of $^{22}$C($I^\pi=0^+$) calculated using the code {\sc efaddy} \cite{Efaddy}.
The probabilities associated with the dominant two-neutron components $^{(2S
+1)}\ell_1$(nn) of each wave function are shown (see also Fig.\ \ref{fig1}) as
are the calculated $^{22}$C point-nucleon matter rms radii, $\langle r^2
\rangle_{22}^{1/2}$, the computed three-body binding energy $E_{3B}$ and the
n+$^{20}$C $2s_{1/2}$ virtual state scattering length $a_0$ for each case.
Wave functions k1--k4 include an attractive central three-body force in the
hyper-radial coordinate, given by Eq. (\ref{3bf}), with strength $V_{3B}$=1.6
MeV. For wave function k5, $V_{3B}=0$. \label{tab1}}
\begin{center}
\begin{ruledtabular}
\begin{tabular}{ccccccccccc}
Model&$V^{\ell=0}_0$&$V_{3B}$ & $E_{3B}$&$a_0$ & $^1 S$(nn)&$^1 D$(nn)&
$^1 G$(nn)&$\langle r^2 \rangle_{22}^{1/2}$&$[2s_{1/2}]^2$&$[1d_{3/2}]^2$ \\
&  (MeV)&  (MeV)&   (MeV)&  (fm)& & & & (fm)& &  \\
\hline
k1& 33.5&1.6& $-$0.442&$-$333.3&0.698&0.234&0.043&3.505&0.931&0.028\\
k2& 33.0&1.6& $-$0.294&$-$ 45.5&0.721&0.213&0.042&3.571&0.927&0.029\\
k3& 32.5&1.6& $-$0.163&$-$ 24.4&0.744&0.193&0.042&3.643&0.923&0.029\\
k4& 32.0&1.6& $-$0.046&$-$ 16.1&0.767&0.174&0.041&3.719&0.919&0.029\\
k5& 33.5&0.0& $-$0.134&$-$333.3&0.743&0.191&0.047&3.669&0.939&0.020\\
\end{tabular}
\end{ruledtabular}
\end{center}
\end{table*}

The parameters of our calculated wave functions and the probabilities
for their most important two-neutron configurations are shown in Table
\ref{table1}. These are all dominated by the neutron $[\nu 2s_{1/2}]^2$
configuration. These calculations use a maximum hypermomentum $K_{\rm max}=
45$. The $^{22}$C point-nucleon rms matter radii $\langle r^2 \rangle_{22
}^{1/2}$ are also shown, computed using $\langle r^2 \rangle_{22}=(20/22)
\langle r^2 \rangle_{20}+ \langle \varrho^2 \rangle/22$, where $\langle
\varrho^2 \rangle$ is the mean-squared hyperradius \cite{rms}. The rms
radius of the $^{20}$C core, $\langle r^2 \rangle_{20}^{1/2}$, is taken
to be 2.913 fm, from the neutron and proton point-particle densities of a
Skyrme Hartree-Fock (HF) calculation and the SkX interaction \cite{Bro98}.
Given the increased mass of the $^{20}$C core in this case, the variations
in the theoretical rms radii obtained ($\approx$6\%) are far more restricted
than was obtained e.g. for $^{11}$Li ($\approx$30\%), as were shown in
Fig. 4 of Ref. \cite{jak2}. We note that none of these $[\nu 2s_{1/2}]^2$
dominated wave functions, that have the correct large-distance three-body
asymptotics, has an rms radius within the 1$\sigma$-error bar on the very
large rms radius, 5.4(9) fm, reported from a simplified reaction description
of a low-energy $\sigma_R$ measurement on a proton target \cite{Tanaka}.

A different set of potential choices was used in Ref.\ \cite{japan}, in which
the unbound $s_{1/2}$ virtual state was also placed close to the $^{21}$C
threshold, while the $1d_{3/2}$ state was located at a significantly higher
energy. In Ref.\ \cite{yamas}, similarly to here, the position of the
$s_{1/2}$ virtual state was varied (there between 0 and 100 keV), however a
very simple contact ($\delta$-function) n-n interaction was used, thus
acting only in n-n relative $s$-waves. Our set of $^{22}$C wave functions
includes systems with $S_{2n}\approx 70$ keV and less, as was suggested
following a recent experimental $^{21}$C search and the use of a zero-range
three-body model analysis by Mosby {\em et al.} \cite{Mos}.

\section{Four-body reaction approaches}

We now describe both the coupled channels and eikonal model methodologies
used for the four-body calculations of the elastic scattering S-matrix and
the reaction cross sections of projectiles (e.g. $^{22}$C), described by
three-body model wave functions, incident upon a nuclear target. We use
and generalise the adiabatic four-body approach as has been discussed
previously by Christley {\it et al.} \cite{Christley}, applied there to
study $^{11}$Li elastic scattering, and also the eikonal model, as was
discussed previously for elastic scattering in Ref.\ \cite{att} and for
reaction cross sections in Refs.\ \cite{jak1,jak2} and elsewhere. We do
not consider the role of Coulomb dissociation, given our interest in light
target nuclei and intermediate energies. The Coulomb interaction, $V_C(R)$,
can be included between the centres of mass of the projectile and target.

\subsection{Adiabatic four-body model calculations}

A detailed formulation of the adiabatic four-body model, as applied to the
elastic scattering of a three-body projectile, was presented by Christley
{\em et al.} \cite{Christley}. In this work, as there, the three projectile
constituents (the neutrons and $^{20}$C core) are assumed to interact with the
target nucleus through complex, spin-independent optical potentials, with the
result that the total spin of the two neutrons, $[s_1 s_2]S$, is a constant
of the motion. In addition to the three-body quantum numbers defined in Figure
\ref{fig1}, we denote the relative orbital angular momentum between the centers
of mass of the composite projectile and the target by $L$ and their separation
by $\vec{R}$.

The adiabatic approximation is common to both the coupled channels and the few-body
Glauber approaches. It assumes that the projectile ($^{22}$C) breakup is predominantly
to low-energy states in the continuum, and thus that the typical energies associated
with the $^{20}$C+n+n three-body Hamiltonian $h(\vec{\rho},\vec{r})$ are small. In
practice one assumes it is a good approximation to replace $h(\vec{\rho},\vec{r})$
by the projectile ground-state energy, $h(\vec{\rho},\vec{r}) \rightarrow E_{3B}$,
in the four-body, $^{22}$C+target Schr\"odinger equation for $\Psi(\vec{R},\vec{\rho},
\vec{r})$. This energy choice, $E_{3B}$, ensures that the dominant elastic components
of the wave function have the correct channel energy and asymptotic properties. Thus,
the adiabatic model solution of the four-body scattering problem, $\Psi^{\rm Ad}(\vec{R},
\vec{\rho},\vec{r})$, satisfies
\begin{equation}
\left[E- E_{3B} - T_R -V_C(R) \right] \Psi^{\rm Ad} = V(\vec{R},\vec{\rho},\vec{r})\,
\Psi^{\rm Ad}\ ,
\label{ad_eqn}
\end{equation}
in which the coordinates $\vec{\rho}$ and $\vec{r}$ that describe the $^{20}$C+n+n
relative motions enter only as parameters and not dynamically. Here $V(\vec{R},
\vec{\rho},\vec{r}) = V_{n1}+V_{n2}+V_{20}$ is the total interaction between
$^{22}$C and the target nucleus, i.e. the sum of the two neutron-target and the
core-target two-body interactions for a given $\vec{\rho}$ and $\vec{r}$. For
calculations of elastic scattering and/or reaction cross sections, i.e of the
$^{22}$C-target elastic S-matrix as a function of $L$, Eq.\ (\ref{ad_eqn}) must
be partial-wave decomposed and solved for all values of $\rho$ and $r$ relevant
to the description of the $^{22}$C ground-state wave function. As we discuss
weakly-bound projectiles, this will involve an extended region of
configuration space.

So, for each fixed radial configuration $({\rho},{r})$ between the neutrons
and the core the collision and excitation of the system is computed by the
solution of an adiabatic (single, fixed energy) radial coupled channels set.
These coupled radial wave functions, $\chi^J_{\alpha'\alpha}(R,\rho,r)$,
involve a chosen set of internal and orbital configurations $\alpha=\{\beta,L\}$,
with $\beta=[\ell_1 \ell_2]{\cal L}$. The coupled equations, with total orbital
angular momentum $J$, with couplings $[{\cal L} L]J$, are
\begin{widetext}
\begin{equation*}
\left[(E-E_{3B}) + \frac{\hbar^2}{2 \mu} \left(\frac{\dd^2}{\dd R^2} - \frac{L'
(L' +1)}{R^2} \right) - V_C(R) \right]\, {\chi}^J_{\alpha'\alpha}(R,\rho,r) = \\
\sum_{\alpha''} V^J_{\alpha'\alpha''}(R,\rho,r)\, {\chi}^J_{\alpha''\alpha}(R,
\rho,r).
\end{equation*}
\end{widetext}
Here, the radial coupling interactions are $V^J_{\alpha'\alpha''}(R,\rho,r) \equiv
\langle \alpha';J| V(\vec{R},\vec{\rho},\vec{r})|\alpha'';J\rangle$, where the
bra-ket notation denotes integration over the angular coordinates of $\vec{R},\
\vec{\rho}$ and $\vec{r}$; see also. Eq. (13) of Ref.\ \cite{Christley}.

The major complication in the computation of these radial couplings is the accurate
treatment of the neutron-target interactions. These terms are genuinely four-body-like
in that they involve and couple the three coordinates $\vec{R},\ \vec{\rho}$ and
$\vec{r}$. We use the generalised multipole expansion of \cite{Christley}. Since the
coupled equations solutions are needed for $\rho$ and $r$ values spanning the entire
$^{22}$C ground-state, we note that accurate calculations of the multipole coupling
potentials, $V_{k_R k_\rho k_r}(R,\rho,r)$, require the use of different techniques
for small values (Gaussian expansion) and larger values (direct numerical integration)
of $r$ and $\rho$ (readers are referred to Appendix A of Ref.\ \cite{Christley}).
The result, from the solution of this set of $R$-dependent coupled channels equations,
is an $(\rho,r)$-dependent reaction matrix, $M^J_{\alpha';\alpha} ({\rho},{r})$, for
the amplitudes of the outgoing waves in each included final-state configuration
$\alpha'$, as computed from the asymptotic ($R\rightarrow \infty$) forms of the
coupled channels radial functions
\begin{equation*}
{\chi}^J_{\alpha'\alpha}(R,\rho,r) \rightarrow F_L(KR)\,\delta_{\alpha' \alpha} +
M^J_{\alpha';\alpha}(\rho,r)\,H^+_{L'}(KR).
\end{equation*}

The physical elastic or breakup amplitudes are now determined by evaluating the
integrals over $\rho$ and $r$ of these adiabatic, $(\rho,r)$-dependent amplitudes
$M^J_{\alpha';\alpha}$ weighted by the probability amplitudes for finding the
$^{20}$C+n+n system with each given $(\rho,r)$ in the entrance and exit channels,
i.e. by the radial wave functions $U_{[\beta S]I}(\rho,r)$. That is,
\begin{eqnarray*}\label{eamp}
{\cal M}^{SJI'I}_{\alpha';\alpha}= \int\!{d\rho}\! \int\!{dr}\, U^{*}_{[\beta'S]I'}
(\rho,r) M^{J}_{\alpha';\alpha}(\rho,r)U_{[\beta S]I}(\rho,r).
\end{eqnarray*}
The relationship of these ${\cal M}$ to the usual partial wave transition amplitudes,
where the physical total angular momentum of the system is ${\cal J}$, having
couplings $[J\,S]{\cal J}$ and $[L\,I]{\cal J}$, is then, with $\hat{k}=\sqrt{2k+1}$
etc.,
\begin{widetext}
\begin{eqnarray}\label{t1}
T^{\cal J}_{L'I' ; LI} = \sum_{\ell_1\ell_2{\cal L}\,\ell'_1\ell'_2{\cal L}'
SJk}(-)^\phi\hat{k}\hat{J}\hat{I}\hat{I'}\,W({\cal L}L{\cal L}'L',Jk)\,W(I
{\cal L}I'{\cal L}^{'},Sk)\,W(LL'II',{k\cal J})\,{\cal M}^{SJI'I}_{\alpha';
\alpha},\label{gen}
\end{eqnarray}
\end{widetext}
where the phase is $\phi=L+L'-J+S+I-I'+{\cal L}+{\cal L}'-{\cal J}-k$. The
elastic scattering amplitudes have $I=I'$ and involve only those channels,
$[[\ell_1 \ell_2]\LL S]I$, that enter the projectile ground state wave
function, Eq.\ (\ref{ground}).

This general angular momentum coupling structure simplifies significantly
when discussing the elastic T-matrix for $^{22}$C scattering, with $I^\pi=0^+$.
The Racah coefficients in Eq.\ (\ref{gen}) then require $k=0$, ${\cal J}=L=L'$,
and ${\cal L}={\cal L}' = S$ for all of the included ground state contributions.
Thus, the T-matrix elements reduce to simplified sums over the most important
ground state components. Defining $T^{\cal J} \equiv T^{\cal J}_{{\cal J}0 ;{
\cal J}0}$  then
\begin{equation}
T^{\cal J} = \sum_{\ell_1 \ell_2
\ell'_1 \ell'_2 J S} \left(\frac{\hat{J}}{\hat{S}
\hat{\cal J}}\right)^{\!2}\,\, {\cal M}^{SJ00}_{\{[\ell'_1 \ell'_2]{S},{\cal J}\};
\{[\ell_1 \ell_2 ]{S},{\cal J}\}} ~.\label{eamp2}
\end{equation}
From these one obtains the elastic S-matrix elements, $S_{22}^{\cal J}=1+2iT^
{\cal J}$, and hence the coupled-channels model reaction cross sections
\begin{eqnarray}\label{cross}
\sigma_R = \frac {\pi}{K^2}\sum_{{\cal J}}\hat{\cal J}^2\,\left(1-| S_{22}^
{\cal J}|^2 \right)\ .
\end{eqnarray}

\subsection{Eikonal four-body model calculations\label{Glaub}}

The four-body adiabatic model Schr\"odinger equation, Eq. (\ref{ad_eqn}), can also
be solved more simply, but more approximately, by use of the eikonal/Glauber
approximation -- that the scattering of the projectile and its constituents is
forward focussed and involves negligible longitudinal (beam directional) momentum
transfers. The formalism and applications of this technique to reaction cross section
calculations for weakly-bound three-body projectiles has been presented in Refs.\
\cite{jak1,jak2} and references therein. Here we summarize the essential results
and the major differences from the coupled channels methodology of the previous
subsection.

Now, the (semi-classical) elastic scattering channel S-matrix of the projectile,
expressed as a function of the impact parameter $b$ of its center-of-mass, is more
simply calculated, based on the eikonal elastic S-matrices for the independent
scattering of its constituent neutrons and core nucleus from the target, as
\begin{eqnarray*}
{\cal S}_{22}(b) = \int\! d^2 \vec{\sigma} \int\! d^2 \vec{s}\ \xi (\vec{\sigma},
\vec{s}) \,{\cal S}_{20}(b_{20})\, {\cal S}_{n1}(b_1) \,{\cal S}_{n2} (b_2).
\label{jj3}
\end{eqnarray*}
Here $\vec{\sigma}$ and $\vec{s}$ are the components of the $^{20}$C+n+n relative
coordinates $\vec{\rho}$ and $\vec{r}$ in the impact parameter plane of the
reaction, the plane normal to the incident beam direction. The Glauber approach
thus avoids the requirement for coupled channels solutions and is computationally
very economical. The individual S-matrices ${\cal S}_{n1}\,$(=$\,{\cal S}_{n2}$)
and ${\cal S}_{20}$ should be calculated from the same neutron- and core-target
complex optical interactions, $V_{n}$ and $V_{20}$, used in the coupled channels
calculations.

The projectile ground-state wave function now enters the calculations through the
$z$-direction-integrated probability density of the projectile, $\xi (\vec{\sigma}
,\vec{s})$, sometimes referred to as the projectile thickness function,
\begin{eqnarray}
\xi (\vec{\sigma},\vec{s}) = \int_{-\infty}^{\infty} d\rho_3 \int_{-\infty}^{\infty}
dr_3 \ \langle\,\, \vert \Phi_{00} (\vec{\rho},\vec{r}) \vert^2 \rangle_{\text{
spin}}\  \label{chi}
\end{eqnarray}
where $\Phi_{00}(\vec{\rho},\vec{r})$ is the $^{22}$C ground state wave function
(with $IM_I=00$) and $\rho_3$ and $r_3$ are the beam direction ($z$-components) of
$\vec{\rho}$ and $\vec{r}$. The relevant $^{20}$C+n+n position probability density
is summed over the neutron spin variables. More explicitly, for the $IM_I=00$ case
of Eq. (\ref{ground}), when ${\cal L}=S$, we are required to calculate
\begin{widetext}
\begin{eqnarray}
\langle \, \vert\Phi_{00} (\vec{\rho},\vec{r}) \vert^2 \rangle_{\text{spin}}
&=& \frac{1}{(4\pi)^2}\  \sum_{\ell_1\ell_1' \ell_2\ell_2'{\cal L}\Lambda}\
(-)^{{\cal L}+\Lambda}\ \hat{\ell_1} \hat{\ell_1'} \hat{\ell_2} \hat{\ell_2'}
\hat{\Lambda}^2\ W(\ell_1\ell_2\ell_1'\ell_2';{\cal L}\Lambda)\,P_{\Lambda}(\cos\gamma)
\nonumber\\ &\times& \left( \begin{array}{rrr} \ell_1 &\ell_1' & \Lambda \\ 0 & 0 &
0 \\ \end{array} \right) \left( \begin{array}{rrr} \ell_2 & \ell_2' & \Lambda
\\ 0 & 0 & 0 \\ \end{array} \right)\ \frac{U_{[[\ell_1\ell_2]{\cal L}{\cal L}]0}(\rho,r)}
{\rho r} \times\frac{U_{[[\ell_1'\ell_2']{\cal L}{\cal L}]0}(\rho,r)}{\rho r}
\end{eqnarray}
\end{widetext}
where $P_{\Lambda}(\cos\gamma)$ is the Legendre Polynomial and $\gamma$ is the
angle between vectors $\vec{\rho}$ and $\vec{r}$. Thus the different angular
momentum components in the projectile ground-state enter explicitly here. The
evaluations of the $z$-directional integrals in Eq.\ (\ref{chi}) for $\xi (
\vec{\sigma},\vec{s})$ use the generalisation of the scheme outlined in Ref.
\cite{att} for the higher orbital angular momentum components $\ell_1$ and
$\ell_2$ that are needed here.

\subsection{Reaction model inputs}

We perform calculations for $^{22}$C incident on a $^{12}$C target at 300
MeV per nucleon. The neutron- and $^{20}$C core-target interactions were
calculated using the single-folding $t_{NN} \rho_t$ model (for the nucleon-target
system) and double-folding $t_{NN}\rho_c \rho_t$ models (for the core-target system).
The inputs needed were the point neutron and proton one-body densities for
the core (c) and target (t) nuclei and an effective nucleon-nucleon (NN)
interaction $t_{NN}$. The $^{20}$C density was taken from a spherical Skyrme
(SkX interaction \cite{Bro98}) HF calculation. The density of the carbon target
 was taken to be of Gaussian form with a point-nucleon root-mean-squared radius
of 2.32 fm. A zero-range NN effective interaction was used. Its strength was
calculated from the free neutron-neutron and neutron-proton cross sections at
the beam energy and the ratio of the real-to-imaginary parts of the forward
scattering NN amplitudes, at 300 MeV, were interpolated (using a polynomial
fit) from the values tabulated from the nucleon-nucleus analysis of Ray
\cite{Ray79}. We note that the use of these folded neutron- and core-target
interactions provided a good description of the recent neutron-removal data
from the heavy carbon isotopes at $~250$ MeV/nucleon \cite{Kob}.  The recent
analysis by Bertulani and De Conti \cite{Carlos} also suggests that further
corrections to this procedure, due to additional Pauli blocking corrections
to the NN effective interaction in the core-target system, should be very small
at the energies of interest here.

Some numerical details of these extended adiabatic model calculations are as
follows. The calculations used an extensively modified version of the code
{\sc ado} developed for the earlier work of Christley {\em et al.} \cite{Christley}.
We use a 24$\times$24 (Gauss-Legendre quadrature) grid of $(\rho,r)$ values on
the interval [0,24] fm, with 12 points on [0,10] fm and 12 points on [10,24] fm
for both $\rho$ and $r$. The time-consuming, four-body neutron-target multipole
coupling potentials $V_{k_R k_\rho k_r}(R,\rho,r)$ are pre-calculated and read
back for each $(\rho,r)$ coupled channels solution. Calculations are carried out
for $R$ values of [0,30] fm (with step 0.005 fm) and 500 partial waves were found
to be adequate for calculations of the entire elastic S-matrix. Calculations where
every partial wave was computed were well reproduced when interpolating the
(smooth) S-matrix every 10th partial wave. The specific breakup configurations in
the coupled channels sets, and projectile ground-state components used, are
discussed in the next section. To allow fair comparisons, calculations using the
Glauber few-body approach used consistent wave function and interactions inputs,
with impact parameters on the range [0,30] fm.

\subsection{Reaction cross section results}

The coupled-channels adiabatic calculations were performed when including several
combinations of ground-state and core+n+n elastic breakup continuum configurations,
$\delta\equiv [[\ell_1 \ell_2]{\cal L}S]I$ (see also Fig.\ \ref{fig1}), to assess
the importance of the different breakup configurations and the convergence of the
calculations. Since the $^{22}$C ground state has spin $I^\pi=0^+$, all ground
state configurations have ${\cal L}=S$ (where $[s_1 s_2]S$) and, in general, $S=0$
or 1. However, the computed three-body wave functions of Table \ref{tab1} actually
contain negligible spin $S=1$ components, in total less that 1.5\% of the full wave
function normalization, and these small fragments are not included. Since, when
assuming central nucleon- and $^{20}$C-target interactions, the $S$ is a good quantum
number of the coupled channels equations, the elastic amplitudes of Eq.\ (\ref{eamp2})
involve only $S=0$ channels.

\begin{figure}[h]
\begin{center}
\includegraphics[angle=-90,width=0.98\figwidth]{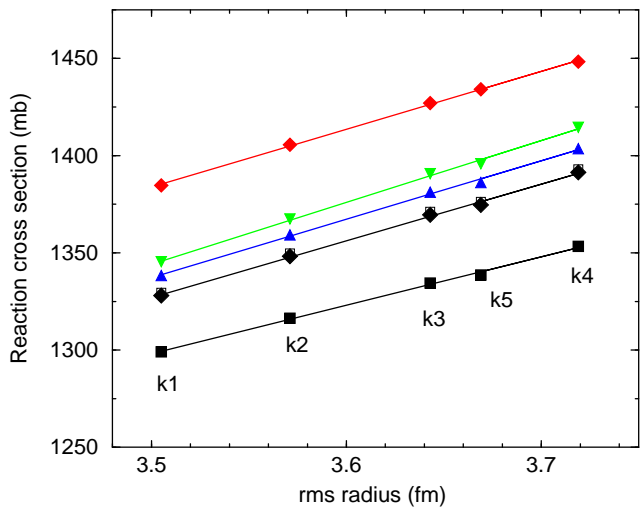}
\end{center}
\caption{\label{fig2} (Color online) Calculated reaction cross sections, for
$^{22}$C on a carbon target at 300 MeV per nucleon, as a function of the rms
radii of the $^{22}$C three-body model wave functions. The different adiabatic
model calculations, $\sigma_R(a)$ through $\sigma_R$(Ad), that systematically
increase the breakup model space are described in the text. The black diamonds
show the results of the most complete adiabatic model calculation $\sigma_R$(Ad)
The black filled squares show the results for the four-body Glauber model
calculations, $\sigma_R$(Gl), for the same three-body wave functions, that
include the [[00]00]0 and [[22]00]0 ground state wave function
components in common with the adiabatic calculations.}
\end{figure}

The $^{22}$C reaction cross sections on a carbon target are calculated for the
different wave functions of subsection \ref{3bwf}, that are summarised in Table
\ref{tab1}. The resulting cross sections are shown in Table \ref{table2} and in
Figure \ref{fig2}, there as a function of the rms radius of the $^{22}$C ground
state. To assess the convergence of the calculations with the breakup model space
different numbers of the $\delta$ configurations are used, shown as different
$\sigma_R(i)$, $i=a,b,c,d$ and Ad, each including successively more (coupled)
angular momentum configurations. Calculations $(a)$ (red diamond symbols) include
only the $[[00]00]0$ configuration, the dominant component (70--77\%, see
Table \ref{tab1}) of the ground-state wave function. Calculations $(b)$ (green
inverted triangles) also include the next most important [[22]00]0 configuration,
that comprises 24--17\% of the ground state. So, calculations $(a)$ and $(b)$
include the major parts of the ground state wave function plus the coupling to
breakup channels with these angular momenta. The next largest component of the
calculated ground state wave functions (4--5\%) has the [[44]00]0 structure, see
below, while a remaining $\approx$1.5\% is distributed over many small $S=1$
fragments, as was discussed above. The importance of this [[44]00]0 ground/breakup
state term is discussed below. Calculations $(c)$, $(d)$ and (Ad) result as we
incrementally expand the breakup model space by adding the couplings to: $(c)$
[[01]10]1 (blue triangles), $(d)$ [[02]20]2 and [[20]20]2 (open squares), and
finally (Ad) the [[22]20]2 (black diamonds) breakup configurations. The latter
two calculations essentially coincide on the figure having cross sections that
differ by less than 2 mb. Figure \ref{fig2} shows the reduced importance of the
channels involving increasing orbital angular momentum transfers among the projectile
constituents. The inclusion of the $\delta=$[[44]00]0 ground-state and breakup
configuration affected the calculated cross sections (for wave functions k1 and
k2) by less than 0.1 mb and was not considered further.

It should be noted that the overall $^{22}$C ground state
wave function was (re)normalized to unity for each subset of the $[[00]00]0$,
[[22]00]0 and [[44]00]0 ground-state configurations included in these results.
We also note that {\em all} of the calculations shown here include projectile
breakup effects, each to a different degree, so, although sharing some visual
similarities, the comparisons shown in Figure \ref{fig2} are different to the
no-breakup versus breakup calculation comparisons made for example in Refs.
\cite{jak1,jak2}.

\begin{table}[ht]
\caption{\label{table2} Calculated reaction cross sections for $^{22}$C,
incident on a carbon target at 300 MeV per nucleon, for the three-body
model wave functions k1--k5 presented in Table \ref{tab1}. The different
intermediate, $(a),(b),(c),(d)$ and final (Ad) adiabatic model and the
Glauber few-body model (Gl) cross sections are discussed in the text.
\label{tab2}}
\begin{center}
\begin{ruledtabular}
\begin{tabular}{lccccc}
Model &k1&k2&k3&k4&k5 \\
$E_{3B}$ (MeV)&$-$0.442&$-$0.294& $-$0.163&$-$0.046&$-$0.134\\
\hline
$\sigma_R(a)$  (mb) &1384.6& 1405.6&1427.0&1448.3&1434.1  \\
$\sigma_R(b)$  (mb) &1345.4& 1367.4&1390.7&1414.5&1395.7  \\
$\sigma_R(c)$  (mb) &1338.3& 1359.2&1381.2&1403.6&1386.1  \\
$\sigma_R(d)$  (mb) &1329.5& 1349.8&1371.1&1392.9&1376.2  \\
$\sigma_R$(Ad) (mb) &1328.0& 1348.3&1369.6&1391.3&1374.6  \\
\hline
$\sigma_R$(Gl) (mb) &1299.1& 1316.3&1334.5&1353.3&1338.6  \\
\end{tabular}
\end{ruledtabular}
\end{center}
\end{table}

Calculations were also carried out using the eikonal (Glauber) few-body approach as
outlined in section \ref{Glaub}. For consistency and fair comparison with the coupled
channels adiabatic approach, these calculations included the same dominant $[[00]00]0$
and [[22]00]0 configurations of the calculated ground state wave functions. The cross
sections obtained are shown in Table \ref{table2}, as $\sigma_R$(Gl), and in Figure
\ref{fig2} (black filled squares). The dependence of the results on the $^{22}$C rms
radius are seen to track those of the coupled channels equations but the magnitudes
of the cross sections are consistently smaller, by $\approx$2\%, than those of
the final, most complete $\sigma_R$(Ad) calculations. This level of agreement
provides a benchmark of the accuracy of the Glauber few-body approach at an energy
of 300 MeV/nucleon, which is valuable given the relative computational efficiency
of the method. We also note that, unlike the coupled channels adiabatic calculations,
there is no explicit truncation of the breakup channels model space used in the
eikonal/Glauber approach, although, based on the observed convergence of the
adiabatic calculations, Figure \ref{fig2}, we do not believe that this is the
source of the small differences observed. The eikonal/Glauber approach also includes
a more approximate treatment of the reaction dynamics, (i) based on the eikonal
elastic S-matrices, ${\cal S}_{n}$ and ${\cal S}_{20}$, that are calculated from
the neutron- and $^{20}$C-target optical potentials assuming straight line paths
through the interaction region, and (ii) assuming there is no longitudinal momentum
transfer in the collisions with the target. Thus, the calculated cross sections show
both the convergence of the adiabatic model calculations and quantify the small
differences between the adiabatic and adiabatic+eikonal dynamical treatments on the
calculated projectile-target absorption $(1-|{\cal S}_{22}|^2)$ and reaction cross
section calculations for the $^{22}$C-target system at the energies of interest here.
These differences may of course be more significant, quantitatively, for other more
exclusive observables and will be clarified in future work in this direction.

\section{Summary}

We have presented converged coupled channels adiabatic four-body reaction model
calculations (using a three-body structure model) of the reaction cross sections
for the last particle-bound carbon isotope $^{22}$C, at intermediate energy.
The coupled channels-based calculations of this work are compared with Glauber
few-body model calculations, that make the additional eikonal approximation in
the solution of the adiabatic Schr\"odinger equation. A set of model three-body
wave functions were obtained, all of which have reasonable two-neutron separation
energies $S_{2n}$, and which have different values for the scattering length
$a_0$ for the assumed virtual $2s_{1/2}$-state in the unbound n+$^{20}$C
two-body subsystem. We have detailed the formulation of the $^{22}$C
elastic scattering S-matrix from which we calculate the reaction cross
sections. The calculations presented include $s-$, $p-$ and $d-$wave
four-body breakup channels. $g$-wave effects were found to be negligible.
The approach used is well-suited to the energy of the present study, 300 MeV
per nucleon, and the calculations reveal the degree of sensitivity of the
reaction cross sections to the assumed structure of the projectile, in
particular its rms radius. Compared to calculations for lighter nuclei,
the range of rms radii from the different $^{22}$C three-body structures is
relatively small with a corresponding smaller cross section sensitivity.
Comparisons of these coupled channels adiabatic results with four-body
Glauber model calculations, based on the same three-body wave functions and
neutron- and $^{20}$C-target interactions, agree to within 2\%.

The approach is relevant to current and future experimental programmes
at fragmentation-based RIB facilities, such as RIBF, FRIB and GSI/FAIR.
The present analysis reveals the high precision of experimental reaction
cross section data that is needed to interrogate the $^{22}$C structure
and the breakup reaction dynamics, pointing to the likely advantage of
more exclusive observables.

\begin{acknowledgments}
This work was supported by a T\"UB\.{I}TAK BIDEB-2219 Postdoctoral Research
Fellowship (for YK) and by the UK Science and Technology Facilities Council
(STFC) (for JAT) under Grant ST/J000051. The authors thank Professor Ian
Thompson and Dr Natalia Timofeyuk for advice on the implementations of the
computer codes {\sc efaddy} and {\sc poler}.
\end{acknowledgments}

\end{document}